\documentclass[aps, pra, twocolumn,notitlepage,nofootinbib]{revtex4-2}
\usepackage{amsmath,amssymb}
\usepackage{graphicx}
\usepackage{newtxtext}
\usepackage{newtxmath}
\usepackage{verbatim}
\usepackage[x11names]{xcolor}
\usepackage[unicode,colorlinks,citecolor=Blue3,linkcolor=Blue3,bookmarks=true]{hyperref}
\usepackage{tikz}
\usepackage{ulem}
\usetikzlibrary {arrows.meta}

\newcommand\bra[2][]{#1\langle {#2} #1|}
\newcommand\ket[2][]{#1|{#2} #1\rangle}

\renewcommand{\Re}{\mathop{\rm Re}\nolimits}
\DeclareMathOperator{\Tr}{Tr}

\begin{document}

\title{Using anti-squeezed Schr\"odinger cat states for detection of a given phase shift}

\author{V.\,L.\,Gorshenin}
\email{valentine.gorshenin@yandex.ru}
\affiliation{Russian Quantum Center, Skolkovo 121205, Russia}
\affiliation{Moscow Institute of Physics and Technology, 141700 Dolgoprudny, Russia}
\author{K.\,D.\,Dyadkin}
\affiliation{Russian Quantum Center, Skolkovo 121205, Russia}
\author{S.\,D.\,Chikalkin}
\affiliation{Russian Quantum Center, Skolkovo 121205, Russia}

\begin{abstract}

We propose to use the antisqueezing-enhanced non-Gaussian Schr\"odinger cat quantum states of the probing light for the task of detection of a given phase shift in optical interferometers. We show that the antisqueezing allows to increase the robustness of the setup to optical losses. We find the optimal degrees of the antisqueezing for experimentally achievable values of the Schr\"odinger cat amplitude and the optical losses and compare the resulting sensitivity with the one provided by the Gaussian squeezed states.
\end{abstract}

\maketitle

\section{Introduction}

At a fundamental level, the phase sensitivity of an interferometer is limited by quantum fluctuations of the probe field and therefore depends on the probe’s quantum state; see, e.g., Refs. ~\cite{Andersen_ch35_2019, 22a1SaKh}. In particular, for a Gaussian coherent squeezed
state, the phase estimation can reach the following value \cite{Caves1981}:
\begin{equation}\label{dphi_sqz}
\Delta\phi_{\rm SQZ} = \frac{e^{-r}}{2\sqrt{N}} \,,
\end{equation}
where \(N\) is the mean number of photons passing through the phase-shifting object(s) and  \(r\) is the logarithmic squeeze parameter (it is assumed here that the squeezing is not very strong, \(e^{2r}\ll N\)).

One may also consider “truly quantum” states, characterized by Wigner quasiprobability functions  \cite{Schleich2001} with a non-Gaussian shape. Their potential metrological utility has been investigated extensively \cite{Holland_PRL_71_1355_1993, Lee_JMO_49_2325_2002, Campos_PRA_68_023810_2003, Berry_PRA_80_052114_2009, Pezze_PRL_110_163604_2013, Perarnau-Llobet_QST_5_025003_2020, Shukla_OptQEl_55-460_2023, Shukla_PhOpen_18_100200_2024, Zheng_PRA_112_3_2025}; see also the review in Ref. \cite{Demkowicz_PIO_60_345_2015}. Overall, these studies suggest that, for phase estimation with no prior information about the phase, non-Gaussian probes offer no substantial advantage. Moreover, the authors of Refs. \cite{Lang_PRL_111_173601_2013, Lang_PRA_90_025802_2014} showed that optimal sensitivity can be attained using comparatively simple, experimentally accessible Gaussian states.

In contrast, non-Gaussian states can be highly effective for another important interferometric task: binary discrimination between two known phase shifts \cite{HelstromBook}. For example, one may wish to discriminate between two dielectrics with known, slightly different refractive indices while depositing less optical energy than would be required when using a coherent
state of light.  This low-energy regime is particularly attractive for interrogating biological specimens that can be damaged by intense illumination \cite{Taylor_NatPhot_7_3_2013, Lotfipour_BiomedOptExpress_16_8_2025}. Without loss of generality, one may set one of the two hypotheses to correspond to zero phase shift, so that the task reduces to detecting the presence of a specified phase shift. 

Note that the general problem of detection an unknown displacement along an unknown axis using of non-Gaussian states was considered in Ref. \cite{Grochowski_PRL_135_23_2025}. 

Note also that “unambiguous state discrimination” refers to an alternative framework introduced in Ref. \cite{Ivanovic_PLA_123_257_1987}; see also Refs. \cite{Enk_S_J_PRA_66_4_2002, Sidhu_J_S_Quantum_7_1025_2023} and the references therein. In this setting, two nonorthogonal states can be discriminated without any error but the protocol succeeds only with some probability $P_D<1$ and produces in the rest of the cases. In many applications, such inconclusive events are equivalent to errors. 

A Schrödinger cat (SC) state is a pure state defined as a superposition of two coherent states \cite{Dodonov_Physica_72_3_1974}. The use of SC states has been proposed for quantum computing \cite{Ralph_PRA_68_4_2003, Ralph_PRL_95_10_2005, Xiang_NatPhot_4_5_2010, Lanyon_NatPhys_5_2_2009, Lund_PRL_100_3_2008}, quantum cryptography \cite{Yin_SciRep_9_1_2019}, and quantum error correction \cite{Schlegel_PRA_106_022431_2022}. 

In Ref. \cite{Gorshenin_LPL_21_6_2024}, we proposed to use the even SC states of the form:

\begin{equation} \label{std_cat}
    \ket{\Psi_0} = \frac{1}{\sqrt{K}}(\ket{\alpha} + \ket{-\alpha})
\end{equation}

for the phase shift detection. Here $\ket{\alpha}$ is a coherent state with the amplitude $\alpha$ and \(K = 2(1+e^{-2|\alpha|^2})\) is the corresponding normalization constant. Throughout, ``the SC state'' refers to the state defined in Eq.~\eqref{std_cat}.

In Ref. \cite{Gorshenin_JOSA_B_42_7_2025}, we considered using Fock states and SC states for this task while accounting for the finite quantum efficiency of photodetection. This inefficiency is equivalent to optical loss and can significantly degrade the observed non-Gaussian properties of the quantum states.

In the phase-space description, optical loss acts as Gaussian smoothing of the Wigner function  \cite{Leonhardt_PRA_48_4598_1993}. Building on this picture, in Ref.~\cite{Filip_PRA_87_4_2013} it was proposed to use antisqueezing to protect the non-Gaussian nature of SC states; there the minimum value of the Wigner function served as the non-Gaussianity measure, and the predicted effect was later verified experimentally \cite{Jeannic_PRL_120_7_2018}. Related ideas were explored in Ref.~\cite{Nugmanov_Arxiv_2201_03257_2022}, which instead quantified non-Gaussianity via the Wigner-negativity volume.

In Ref. \cite{Gorshenin_LPL_21_6_2024}, we introduced a phase-shift detection protocol based on SC states. In a subsequent study \cite{Gorshenin_JOSA_B_42_7_2025}, we showed that even weak optical losses can substantially degrade the performance of this protocol. In this paper, we investigate antisqueezing as a method for protecting the non-Gaussian properties of SC states in fixed phase-shift detection. We quantify this protection through the preservation of the negative volume of the Wigner function. Our analysis includes optical losses both before the interferometer, such as quantum-state-preparation inefficiency, and after the interferometer, such as photodetection inefficiency. We then numerically optimize the antisqueezing strength for optimistic but experimentally realistic values of the Schrödinger cat amplitude and loss. Finally, we demonstrate that, for the fixed phase-shift detection task considered here, antisqueezed SC states can surpass the sensitivity achievable with Gaussian squeezed states.

The paper is organized as follows. Section~\ref{sec:method-description} introduces the interferometric schemes considered in this paper. In Sec.~\ref{sec:losses-on-sc-state-wigner-func} we analyze how the Wigner function of the probe light is affected by the optical losses. In Sec.~\ref{sec:phot-stat} we calculate the photon-number statistics of the output optical state with account for the optical losses. In Sec.~\ref{sec:max-likelihood-est} we discuss the data processing procedure. In Sec.~\ref{sec:estimates}, we calculate the detection errors for a realistic implementation of our scheme. Finally, in Sec.~\ref{sec:discussion} we conclude with the results of the paper.

\section{Optical scheme}\label{sec:method-description}

\begin{figure}
\includegraphics[width=0.85\linewidth]{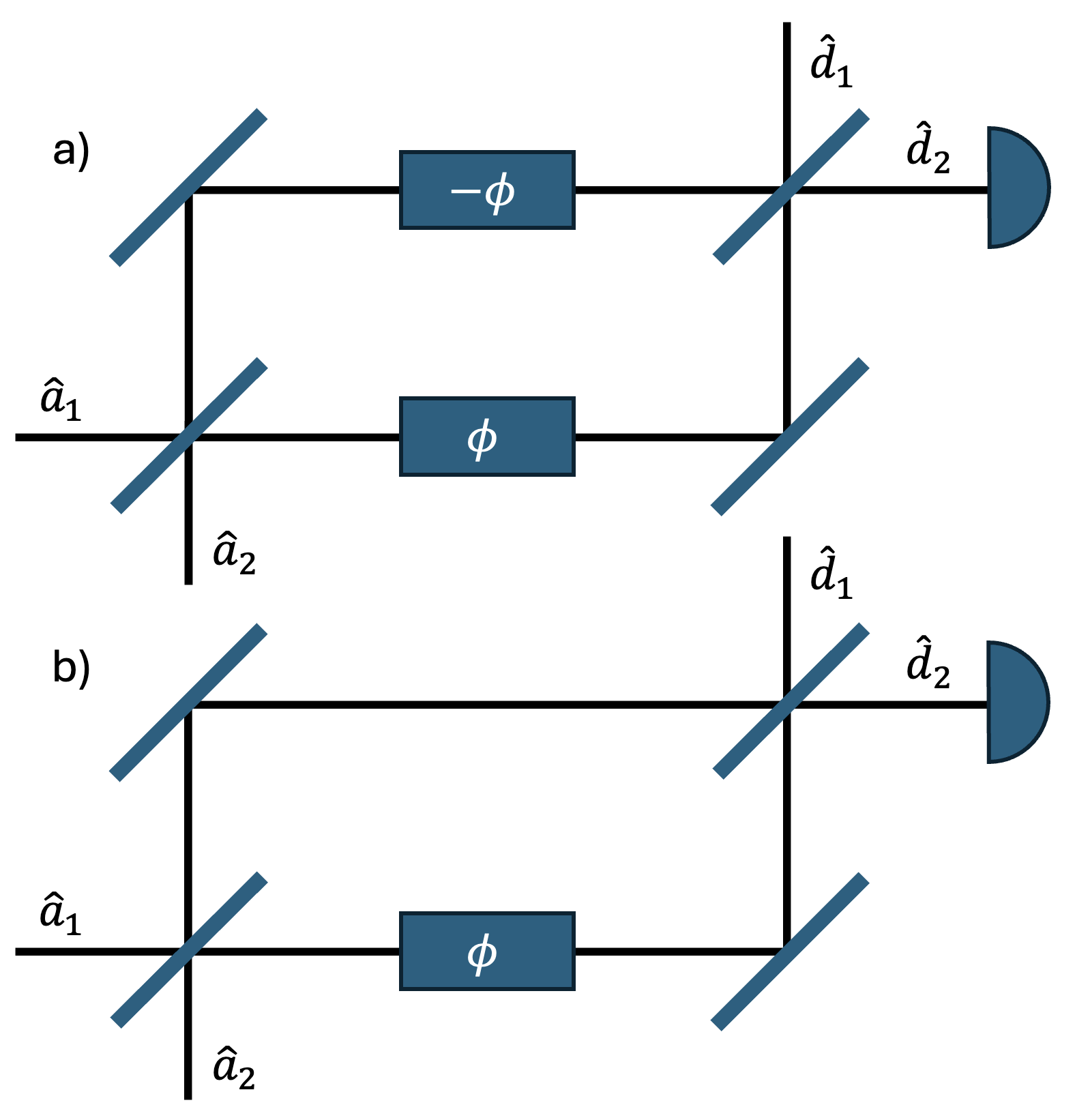}
\caption{
	Two equivalent implementations of the Mach-Zehnder interferometer: antisymmetric (a) and asymmetric (b) ones. \(\hat{a}_1\) and \(\hat{d}_1\) -- bright input and output ports, \(\hat{a}_2\) and \(\hat{d}_2\) -- dark input and output ports.}
\label{fig:scheme}
\end{figure}

Following Ref.~\cite{Gorshenin_LPL_21_6_2024} we consider two standard configurations of the two-arm Mach-Zehnder interferometer (Fig.~\ref{fig:scheme}). In the asymmetric configuration, strongly unbalanced beam splitters are employed and the signal phase shift $\phi$ is imparted to a single arm. In the antisymmetric configuration, equal-magnitude phase shifts of opposite sign $\pm\phi$ are applied to the two arms. 

In both configurations, a classical coherent field enters through the bright input port, and a quantum state is injected through the dark input port. The interferometer is operated in the dark fringe regime, where, at $\phi=0$, the output states coincide with the input states at both the dark and bright ports. 

In the small phase shift $\phi \ll 1$, the bright output port field is insensitive to $\phi$, while the state emerging from the dark port undergoes an effective displacement described by the operator ((see Ref.\,\cite{Gorshenin_LPL_21_6_2024}):

\begin{equation}\label{eq:displacement_oper_definition}
\hat{\mathcal{D}}(i \delta_0) = e^{i\delta_0(\hat{a} + \hat{a}^\dag)} \,,
\end{equation}
where $\hat{a}$ denotes the annihilation operator and \(\delta_0\) is a dimensionless displacement parameter that scales with the bright-port coherent amplitude \(\sqrt{N}\) acting on the phase:

\begin{equation}\label{eq:delta-0-defenition}
\delta_0 = \sqrt{N}\phi \,.
\end{equation}
Here \(N\) is the mean photon number impinging on the phase-shifting element(s).

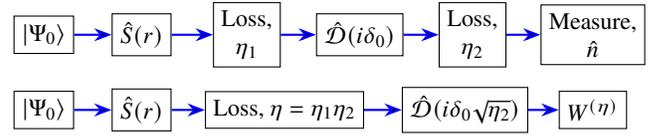
\begin{figure}
	\centering
	\begin{tikzpicture}
		\node[draw, align=center] at (0,0) (A) {\(\ket{\Psi_0}\)};
		\node[draw, align=center] at (1.3,0) (B) {\(\hat{S}(r)\)};
		\node[draw, align=center] at (2.7,0) (C) {Loss, \\ \(\eta_1\)};
		\node[draw, align=center] at (4.2,0) (D) {\(\hat{\mathcal{D}}(i \delta_0)\)};
		\node[draw, align=center] at (5.7,0) (E) {Loss, \\ \(\eta_2\)};
		\node[draw, align=center] at (7.3,0) (F) {Measure, \\ \(\hat{n}\)};
		\draw [draw = blue, thick,
		arrows={
			-Stealth[]}]
		(A) edge (B);
		\draw [draw = blue, thick,
		arrows={
			-Stealth[]}]
		(B) edge (C);
		\draw [draw = blue, thick,
		arrows={
			-Stealth[]}]
		(C) edge (D);
		\draw [draw = blue, thick,
		arrows={
			-Stealth[]}]
		(D) edge (E);
		\draw [draw = blue, thick,
		arrows={
			-Stealth[]}]
		(E) edge (F);

		\node[draw, align=center] at (0,-1) (A) {\(\ket{\Psi_0}\)};
		\node[draw, align=center] at (1.3,-1) (B) {\(\hat{S}(r)\)};
		\node[draw, align=center] at (3.2,-1) (C) {Loss, \(\eta = \eta_1 \eta_2\)};
		\node[draw, align=center] at (5.6,-1) (D) {\(\hat{\mathcal{D}}(i \delta_0 \sqrt{\eta_2})\)};
		\node[draw, align=center] at (7.3,-1) (E) {\(W^{(\eta)}\)};
		\draw [draw = blue, thick,
		arrows={
			-Stealth[]}]
		(A) edge (B);
		\draw [draw = blue, thick,
		arrows={
			-Stealth[]}]
		(B) edge (C);
		\draw [draw = blue, thick,
		arrows={
			-Stealth[]}]
		(C) edge (D);
		\draw [draw = blue, thick,
		arrows={
			-Stealth[]}]
		(D) edge (E);
	\end{tikzpicture}
	\caption{(Top) Diagram of the lossy interferometer. (Bottom) Diagram of the equivalent simplified optical scheme.}
	\label{fig:principal-scheme-two-optical-losses}
\end{figure}

A schematic diagram of the interferometric scheme considered here is shown in Fig.~\ref{fig:principal-scheme-two-optical-losses} (top). We start with the SC state in Eq.~\eqref{std_cat}. We then apply an antisqueezing operation and inject the resulting state into the interferometer. The output state is measured using a photon-number-resolving (PNR) detector. 

We account for optical loss by introducing two ``loss'' blocks in the diagram, placed before and after the interferometer and characterized by quantum efficiencies $\eta_1$ and $\eta_2$, respectively. The first block accounts for losses introduced by the antisqueezer as well as coupling (input) losses of the interferometer. The second block accounts for the interferometer output losses and the finite quantum efficiency of the photodetector. We model optical loss using the standard effective beam splitter model \cite{Leonhardt_PRA_48_4598_1993}.

It is shown in Appendix~\ref{app:two-seq-phot-loss} that this optical scheme is equivalent two the simpler one with the single source of losses located before the interferometer, see Fig.~\ref{fig:principal-scheme-two-optical-losses} (bottom), and with the following effective parameters:
\begin{subequations}
  \begin{gather}
    \text{effective loss factor:}\quad\eta = \eta_1\eta_2 \,, \\
    \text{effective signal:}\quad\delta = \delta_0\sqrt{\eta_2} \,.
  \end{gather}
\end{subequations}
In the Sec.\,\ref{sec:losses-on-sc-state-wigner-func} and Sec.\,\ref{sec:phot-stat} we will use this effective model.

\section{Optical losses} \label{sec:losses-on-sc-state-wigner-func}

The Wigner function of a lossy anti-squeezed SC state has been derived in Ref.~\cite{Jeannic_PRL_120_7_2018}; nevertheless, we present the main result here for completeness and to set the stage for the subsequent discussion. The wave function of the anti-squeezed and subsequently displaced Schr\"odinger-cat state, prior to optical loss, has the form:
\begin{equation}\label{eq:sq-sc-state-def}
	\ket{\psi_{\rm sq,\,cat}} = \frac{1}{\sqrt{2 (1 + \exp(-2|\alpha|^2|)}} \hat{S}(r) \big(\ket{\alpha} + \ket{-\alpha}\big)
\end{equation}
where the squeeze operator \(\hat{S}(r)\) is defined as follows:
\begin{equation}
\hat{S}(r) = \exp\Big(\frac{r}{2} \big(\hat{a}^{2} - \hat{a}^{\dagger 2} \big)\Big)
\end{equation}
Throughout, we use the convention that logarithmic squeeze parameter \(r > 0\) corresponds to anti-squeezing and \(r < 0\) to squeezing.

Wigner function after passing equivalent optical scheme presented on bottom panel of Fig. \ref{fig:scheme} (shown in the Appendix \ref{app:wig-func-calc}) is follows:
\begin{equation}
	W^{(\eta)}(x,p) = F \big(W_{\rm int} + W_{+} + W_{-} \big)
\end{equation}
where the interference part \(W_{\rm int}\) and the Gaussian bell parts \(W_{\pm}\) of Wigner function are equal to
\begin{equation}\label{eq:int_part_sc_wigner}
W_{\rm int} = 2 \exp
\Big(
-C
-B(p-\delta)^2
-A x^2
\Big)
\cos
\Big(
D (p-\delta)
\Big)
\end{equation}
\begin{equation}\label{eq:bell_part_sc_wigner}
W_{\pm} = \exp \Big(
-A (x\pm\xi)^2 - B(p-\delta)^2
\Big)\,,
\end{equation}
where non-negative coefficients \(A, B, C, D\) and \(F\) are defined as follows:
\begin{gather}\label{eq:sc-state-wigner-func-pars}
	A = 2 \frac{s^2}{s^2 (1-\eta) + \eta}, \quad
	B = 2 \frac{1}{1 + (s^2 - 1)\eta},\, \\
	C = 2 \frac{\alpha^2 (1-\eta)}{1+\eta(s^2 - 1)}, \quad
	D = \frac{4 s \alpha \sqrt{\eta}}{1+\eta(s^2 - 1)},\,\\
	\label{eq:sc-state-wigner-after-single-losses}
	F = \frac{s }{\pi  \left(1+e^{-2 \alpha ^2}\right)}\sqrt{\frac{1}{(\eta  (s^2-1)+1) (s^2(1-\eta)+\eta)}}
\end{gather}
where \( s \equiv e^r\). In the lossless and unsqueezed limit (\(\eta = 1,\,s=1\)), they simplify to
\begin{equation}
	A=B=2, \quad C = 0, \quad D = 4 \alpha, \quad
	F=\frac{1}{\pi(1+e^{-2\alpha^2})}
\end{equation}

which reproduces the Wigner function of the SC state in the absence of squeezing and loss [see Eq.~\eqref{eq:cat-state-wigner-func-before-losses}].

A standard quantitative witness of nonclassicality is the Wigner negativity: the integrated ``volume'' of the negative region of the Wigner function. This negativity captures the genuinely quantum aspects of the state that can sharpen distinguishability, and in favorable regimes, drive the overlap toward (near) orthogonality. In this section, we develop an approximate approach for estimating the Wigner-negativity volume of large Schrödinger cat states in the regime of large cat amplitude and pronounced antisqueezing.

To quantify the nonclassicality of the lossy SC state, we use the volume of the negative part of its Wigner function (Wigner negativity) \(V_{\rm neg}\) defined by:

\begin{equation}
	V_{\rm neg} \equiv \frac{1}{2}\int_{-\infty}^\infty dx \int_{-\infty}^\infty dp \left(\big|W^{(\eta)}(x,p) \big|-W^{(\eta)}(x,p)\right)
\end{equation}

This negativity captures the genuinely quantum aspects of the state that can sharpen distinguishability, and in favorable regimes, drive the overlap toward (near) orthogonality. In Appendix~\ref{app:negative-volume-of-sc-stae-derivation}, we develop an approximate approach for estimating the Wigner negativity volume of bright Schrödinger cat states in the regime of large cat amplitude and performed antisqueezing, which gives

\begin{equation} \label{eq:negative-volume-sc-state-sq-after-losses-approx}
	V_{\rm neg} =
	\frac{e^{-\frac{2 \alpha ^2 (1-\eta )}{\eta \left(s^2-1\right)+1}} }{\pi  \left(e^{-2 \alpha ^2}+1\right)}
	\vartheta_3\left(\frac{\pi }{2},e^{-\frac{2 s^2 \alpha ^2 \eta }{\eta(s^2 - 1)+1}}\right)
    ,
\end{equation}
where \(\vartheta_3(x)\) is the Jacobi theta-function.

\begin{figure}
	\centering
	\includegraphics[width=0.86\linewidth]{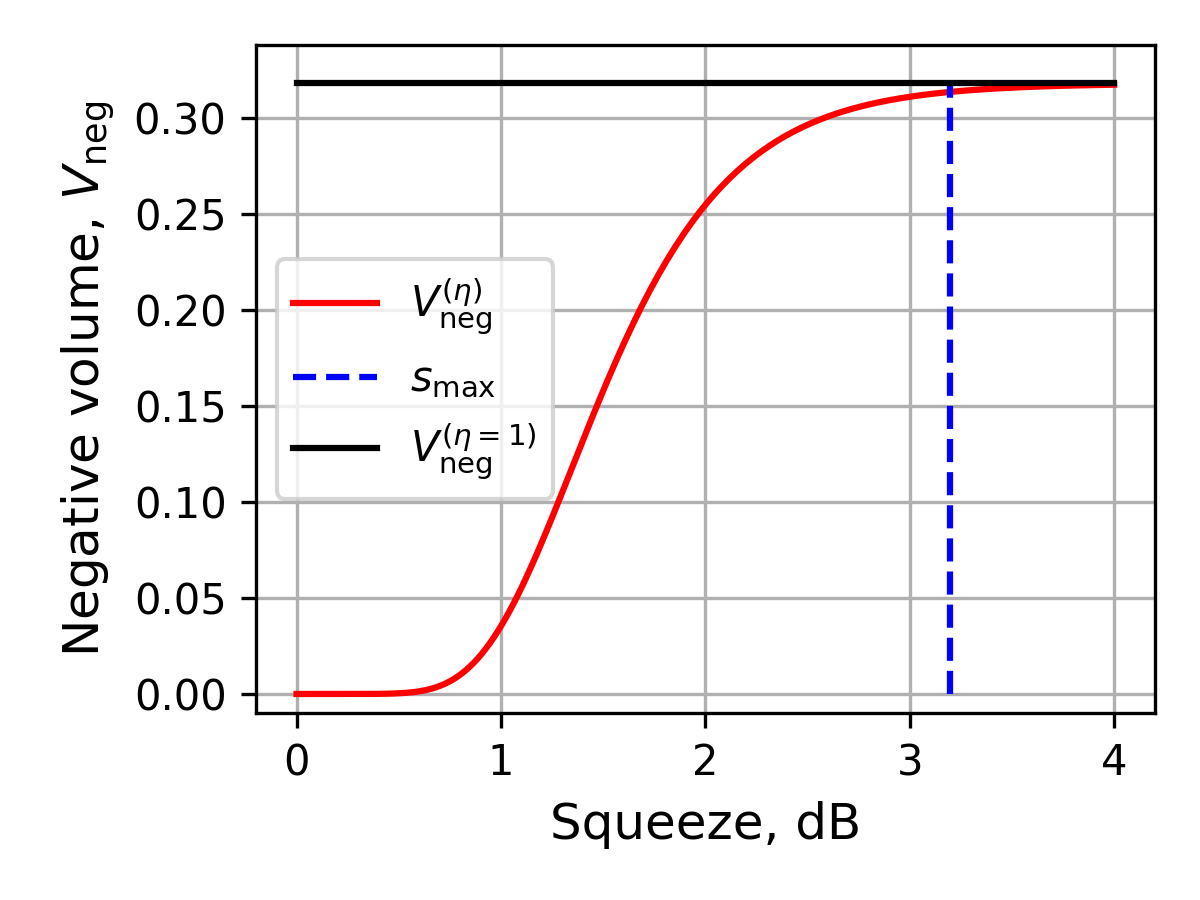}
	\caption{
		Wigner negativity volume \(V_{\rm neg}\) of an antisqueezed SC state versus the antisqueezing level (in dB). Red solid curve: \(\eta = 0.9\); black solid line: \(\eta = 1\). In both cases, \(\alpha = 10\). The blue vertical dashed line marks the validity threshold of the approximation \(s_{\rm max}\), see Eq.~\eqref{eq:v-neg-approx-eq-applicability-condition}.
	}
	\label{fig:neg-volume-of-wigner-func}
\end{figure}

Figure~\ref{fig:neg-volume-of-wigner-func} plots the Wigner-negativity volume after loss versus the antisqueezing factor \(s\), for representative values of the SC state amplitude \(\alpha\) and the loss parameter \(\eta\). Notably, already at \(\sim 3\) dB of antisqueezing the Wigner negativity remains sizable, indicating that antisqueezing can substantially mitigate loss-induced degradation.

\section{Photon-number statistics of output state after optical losses}\label{sec:phot-stat}
To incorporate the effect of optical loss in the output state into the phase-shift detection procedure, we need to calculate the photon-number probabilities \(p_{n}\) of the resulting state. We denote the corresponding density operator by \(\hat{\rho}\). These probabilities can be obtained conveniently from a standard property of the Wigner function \cite{Schleich2001}:

\begin{equation}\label{eq:prob-fock-space-property}
p_n \equiv \Tr \big(\ket{n}\bra{n}\hat{\rho}\big)=
2 \pi
\int dx
\int dp
W_\rho(x,p) W_n(x,p)\,,
\end{equation}
where $W_n(x,p)$ is the Wigner function of the Fock state. The photon statistics of the resultant state is as follows (see for details Appendix~\ref{app:phot-stat}):

\begin{equation}\label{eq:prob-sc-loss-fine}
	\begin{aligned}
		p_n
		=
		4 \pi
		\sum_{k=0}^{k=n} \sum_{m=0}^{m=k}
		\frac{n!}{k! (n-k)!}
		\frac{\sqrt{\pi } F 2^{2 k+1}  (-1)^{k+n} }
		{(A+2)^{m+\frac{1}{2}}
			(B+2)^{k-m+\frac{1}{2}}}
		\\
		\exp \left(-\frac{2 A \xi^2}{A+2}-\frac{D^2}{4 B+8}-\frac{2 B \delta
			^2}{B+2}-C\right)
		\cdot\\\cdot
		\Big(\sqrt{\pi } e^{\frac{D^2}{4 B+8}+C} L_m^{-\frac{1}{2}}\left(-\frac{A^2 \xi^2}{A+2}\right) L_{k-m}^{-\frac{1}{2}}\left(-\frac{B^2 \delta^2}{B+2}\right)
		+\\+
		e^{\frac{2 A \xi ^2}{A+2}} \frac{\Gamma\left(m+\frac{1}{2}\right) }{m!}
		\Re\left(
		e^{\frac{2 i D \delta }{B+2}}
		L_{k-m}^{-\frac{1}{2}}
		\left(\frac{(D+2 i B \delta )^2}{4(B+2)}\right)
		\right)\Big)\,.
	\end{aligned}
\end{equation}

Numerical evaluation of the ratio \(\Gamma(m+1/2)/m!\) slightly complicates the computation of the photon-number statistics. Using Stirling’s approximation, we derive a simpler expression that is accurate and numerically stable; see the end of Appendix~\ref{app:phot-stat}.

The alternative approach to the calculation of the photon number statistics is the use of the binomial conditional photon number probabilities introduced by the loss sources. However, this method does not readily accommodate sequences of the form loss \(\rightarrow\) linear operation (displacement, squeezing, and phase shift) \(\rightarrow\) loss. Therefore, to validate our approximate approach, in Appendix~\ref{app:numerical-calc-comparison} we compare both methods in a simpler setting: a displaced and squeezed SC state followed by a single source of losses, and show that their results coincide with very good precision.

\section{Data processing algorithm}\label{sec:max-likelihood-est}

In a single-shot setting, the decision must be made from a single observed photon count \(n\) rather than from an estimated distribution. In Ref.~\cite{Gorshenin_LPL_21_6_2024}, a simple decision strategy for determining the presence or absence of a phase shift based on photon-number parity was used. However, it becomes suboptimal in the presence of optical loss. In Ref.~\cite{Gorshenin_JOSA_B_42_7_2025}, we proposed using the maximum-likelihood method. We use this method here as well.

We assign each measured photon number to one of the two hypotheses: the absence or presence of a phase shift \cite{Helstrom_InformationAndControl_10_3_1967, HelstromBook}. For binary discrimination between the non-displaced state \(\hat{\rho}_0\) and the displaced state \(\hat{\rho}_\delta\), we consider their photon-number distributions

\begin{equation}
    p_n^{(0)} \equiv \Tr (\hat{\rho}_0 \ket{n}\bra{n})
    \,\quad
    p_n^{(\delta)} \equiv \Tr (\hat{\rho}_\delta \ket{n}\bra{n})\,.
\end{equation}

We therefore partition the non-negative integers into two decision regions, \(N_0\) and \(N_\delta\) associated with the hypotheses ``phase-shift absence'' and ``phase-shift present'', respectively. For a single-shot defined decision rule, these regions must form a complete partition of the Fock-state index set. We assign n according to the maximum-likelihood (ML) decision rule:

\begin{equation} \label{eq:threshold-defenition}
	n \in N_0 \iff p_n^{(0)} >p_n^{(\delta)}
	,\,\quad
	n \in N_\delta \iff p_n^{(0)} < p_n^{(\delta)}\,.
\end{equation}

We define the false-positive and false-negative error probabilities, \(p_{\rm fp}\) and \(p_{\rm fn}\), as follows:
\begin{equation} \label{eq:prob-fp-and-fn-def}
    p_{\rm fp} = \sum_{n\in N_\delta} p_n^{(0)}
    ,\quad
    p_{\rm fn} = \sum_{n\in N_0} p_n^{(\delta)}\,.
\end{equation}
The corresponding total error is defined as:
\begin{equation}
  p_{\rm tot} \equiv p_{\rm fn} + p_{\rm fp} \,.
\end{equation}

\section{Estimates of the error probabilities}\label{sec:estimates}

We proceed to estimate the false-positive and false-negative error probabilities, adopting reasonably optimistic parameter values for our scheme. Because the method introduced in Sec.~\ref{sec:phot-stat} becomes computationally intractable for photon number higher than 23 in the following numerical calculation realization, we instead employ a binomial method for loss consideration. This restriction precludes the inclusion of SC states with large amplitudes. We consider displacement with argument \(i \delta_0\) and a realistic detector with non-unity quantum efficiency. 

Concerning the quantum efficiency $\eta$, over the past several decades, considerable progress has been made in developing PNR detectors. The best values of \(\eta\) are currently obtained with cryogenic photon-counting technologies, most notably superconducting nanowire detectors and transition-edge sensors (TES). The former ones can reach efficiencies as high as 95\% while resolving on the order of 20 photons \cite{Stasi_PRAppl_19_064041_2023}. TES-based PNR detectors can achieve efficiencies up to 98\% and resolve up to $\sim 100$ photons \cite{Lita_OE_16_3032_2008, Fukuda_OE_19_870_2011, Gerrits_OE_20_23798_2012}.

Regarding the input quantum state preparation, experimentally demonstrated SC states preparation protocols typically achieve amplitudes  \(\alpha \le 1.5\) \cite{Ourjoumtsev_Science_312_83_2006, Huang_PRL_115_023602_2015, Sychev_NatPhot_11_379_2017, Jeannic_PRL_120_7_2018}. Looking ahead, PNR-detectors-based schemes for generating brighter SC states have been proposed, with target amplitudes on the order of \(\alpha \approx 4-5\) \cite{Kuts_PhyicaScripta_97_11_2022, Podoshvedov_SR_13_3965_2023}.

The values of squeezing, experimentally attainable in non-linear crystals, is typically in the range \(5-10\) dB, which corresponds to to \(r\) in the range \( 0.57 \leq r \leq 1.15\) \cite{Inoue_ApplPhysLett_122_104001_2023, Iskhakov_OptLett_41_10_2016, Frascella_NPJQuantInform_7_1_72_2021, Jeannic_PRL_120_7_2018}.

\begin{figure}
\centering
\includegraphics[width=0.86\linewidth]{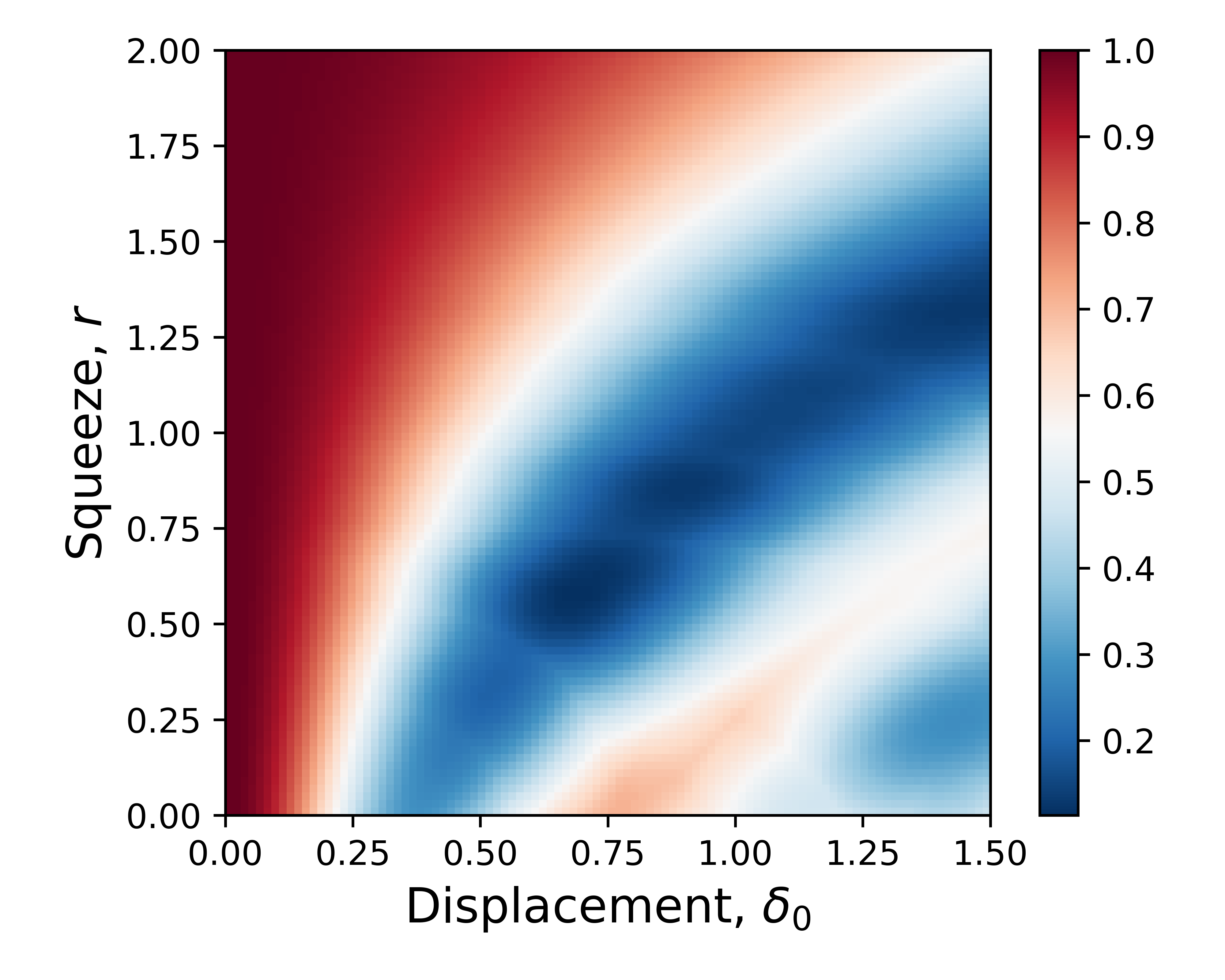}
\caption{
	Total error probability \(p_{\rm tot}\) as a function of the displacement \(\delta_0\) and antisqueezing factor \(r\) for \(\alpha = 2.0\) and \(\eta = 0.975\).
}
\label{fig:tot-error-sq-and-displ-variate}
\end{figure}

In Fig.~\ref{fig:tot-error-sq-and-displ-variate}, the detection error $p_{\rm tot}$ is plotted as a function of $\delta_0$ and $r$ for the particular case of \(\alpha = 2\) and \(\eta = 0.975\). This plots shows several minima of the total detection error $p_{\rm tot}$. The best value (for these specific parameters) is close to $p_{\rm tot}\approx0.1$ and can be reached using moderate antisqueezing of $\approx 5\,{\rm dB}$ at $\delta_0\approx0.7$.

\begin{figure}
\centering
\includegraphics[width=0.86\linewidth]{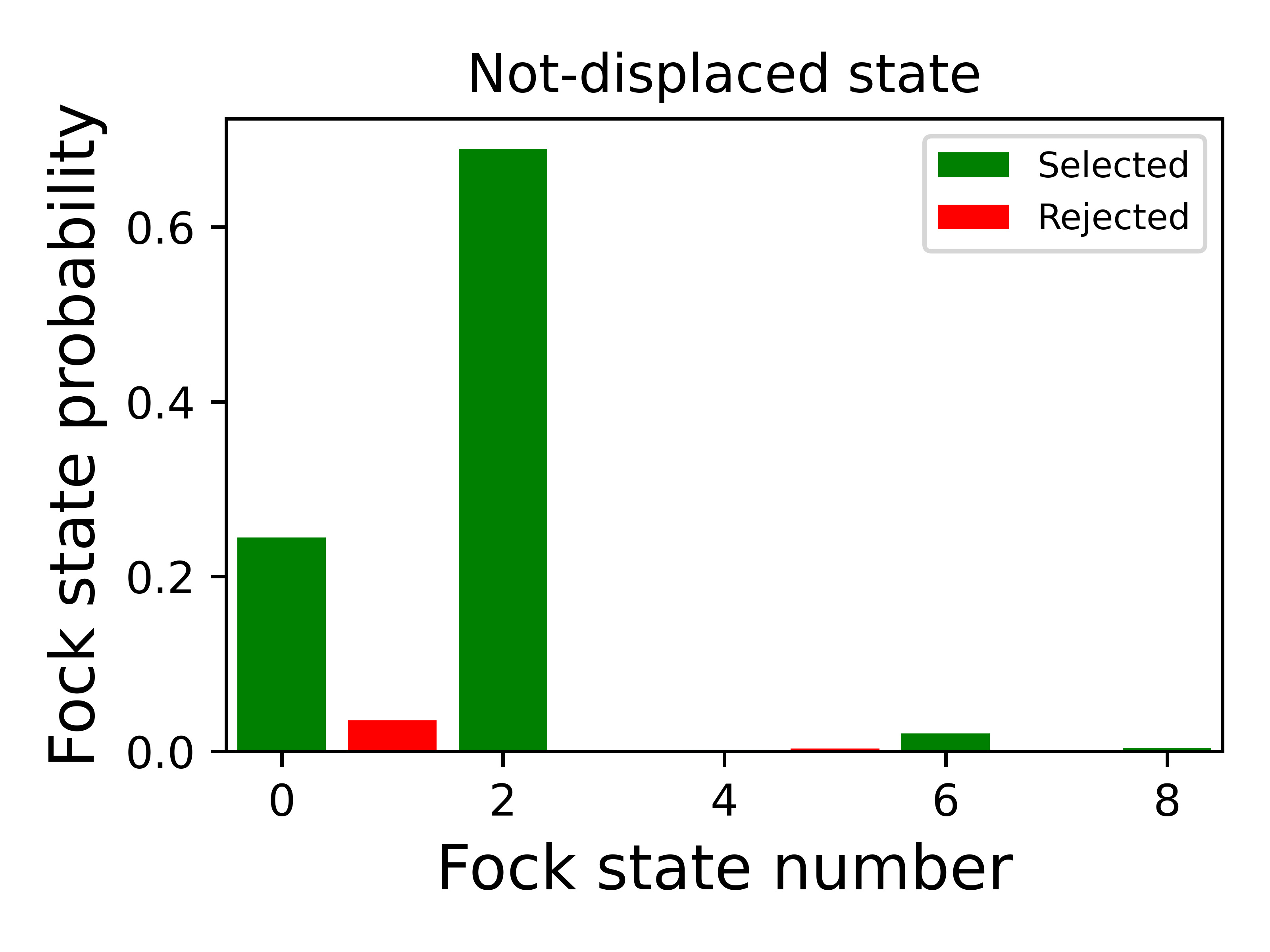}
\includegraphics[width=0.86\linewidth]{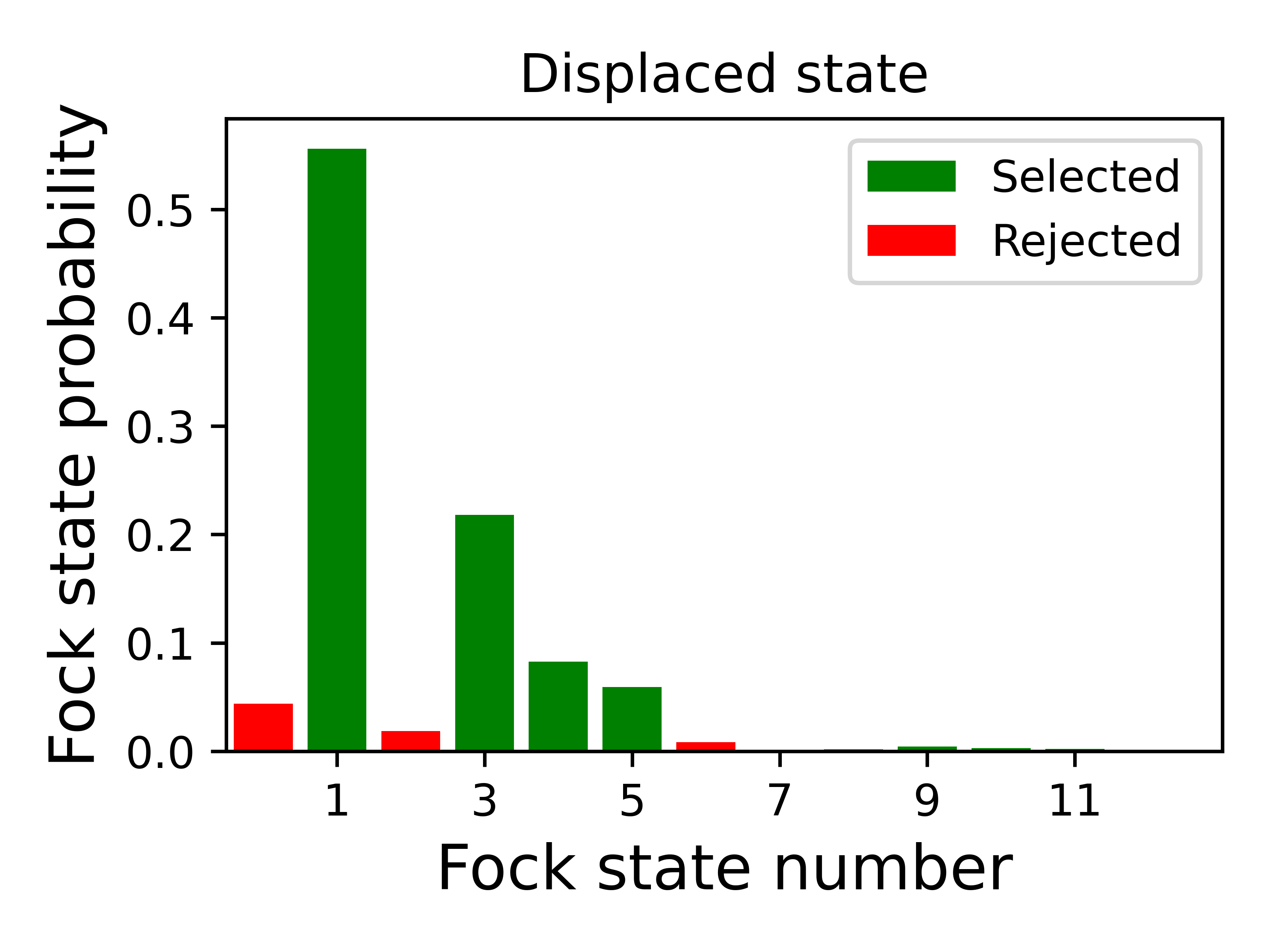}
\caption{
	Photon-number distributions of the optimized anti-squeezed SC state for the case of \(\alpha = 2.0\) and \(\eta = 0.975\), $r=0.56$ ($5\,{\rm dB}$). Top: no signal, $\delta_0=0$, bottom: with signal, $\delta_0=0.68$. The red bars constitute the detection error.
}
\label{fig:primitive-vs-selective-appr}
\end{figure}

The corresponding photon-number distributions of the SC state (after losses) for the cases of the absence and the presence of the signal are plotted in Fig.~\ref{fig:primitive-vs-selective-appr}.

\begin{figure}
	\centering
	\includegraphics[width=0.93\linewidth]{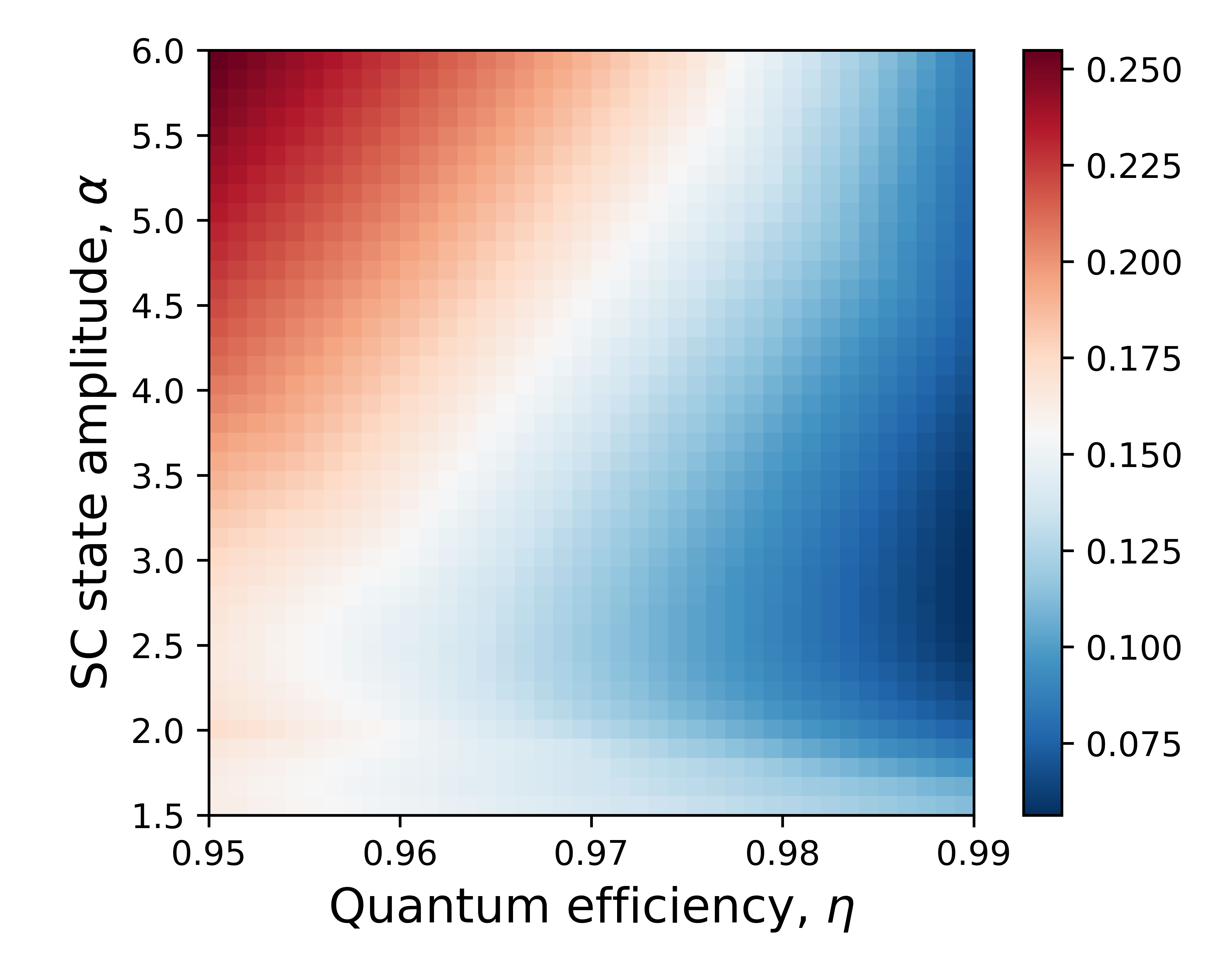}
	\caption{
		Total error probability \(p_{\rm tot}\)  as a function of the SC state amplitude \(\alpha\) and the quantum efficiency \(\eta\). For each \((\alpha, \eta)\), the displacement \(\delta_0\) and antisqueezing parameter \(r\) are optimized.
	}
	\label{fig:by-ampl-and-loss-p-tot}
\end{figure}

A more general view on the sensitivity is provided by Fig.~\ref{fig:by-ampl-and-loss-p-tot}, where \(p_{\rm tot}\) is plotted as a function of the quantum efficiency \(\eta\) and the SC amplitude \(\alpha\). For each pair of these arguments, the global minimum of \(p_{\rm tot}\) in $(\delta_0,r)$ is calculated and presented in the plot.

\begin{figure}
	\centering
	\includegraphics[width=0.93\linewidth]{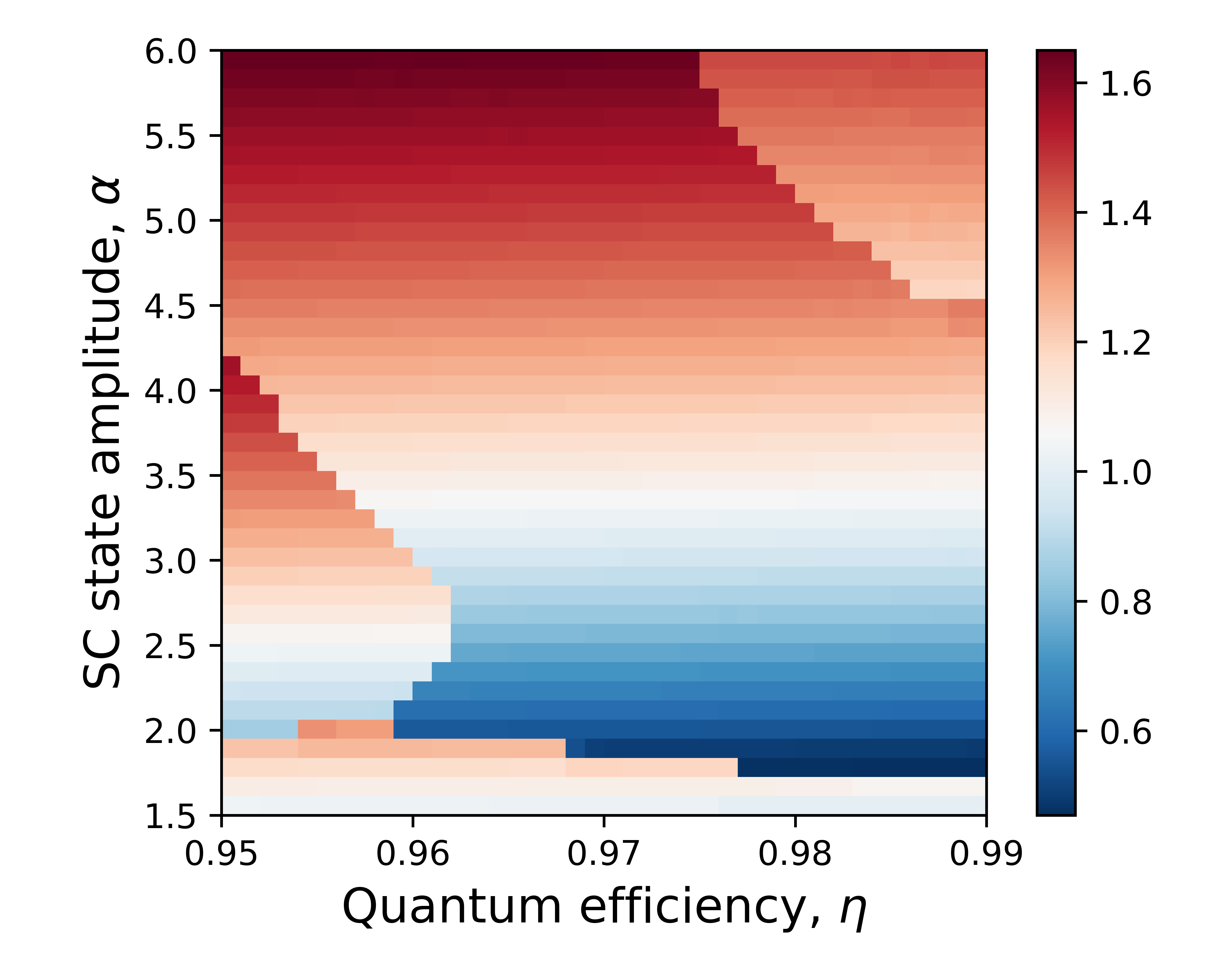}
	\caption{
		The optimal antisqueezing \(r\) as a function of the SC state amplitude \(\alpha\) and the quantum efficiency \(\eta\).
	}
	\label{fig:by-ampl-and-loss-r}
\end{figure}

In Fig.~\ref{fig:by-ampl-and-loss-r}, the corresponding optimized factor r is plotted as a function of $\eta$ and $\alpha$, showing that the required antisqueezing tends to decrease with the increase of $\eta$ and decreasing \(\alpha\), for the considered here values of $\eta$ and $\alpha$, never exceeds $\approx14{\rm dB}$.

\begin{figure}
	\centering
	\includegraphics[width=0.93\linewidth]{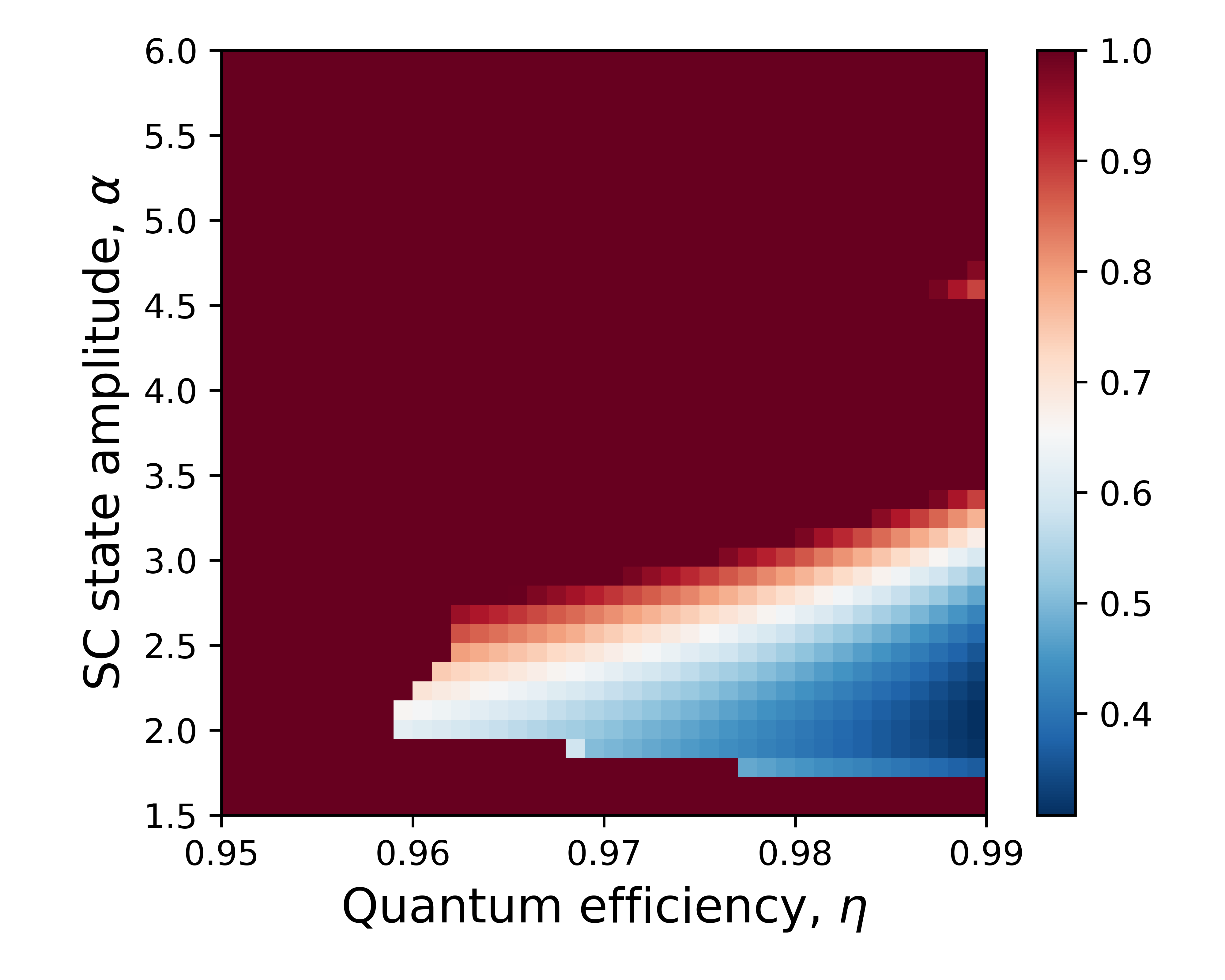}
	\caption{
		The ratio of the detection errors for the SC state and Gaussian squeezed state \(p_{\rm tot}/p_{\rm sq}\) as a function of $\alpha$ and $\eta$. For each \((\alpha, \eta)\), the displacement \(\delta_0\) and antisqueezing parameter \(r\) are optimized. To emphasize the regime in which the SC state beats the squeezed vacuum one, the values of \(p_{\rm tot}/p_{\rm sq} > 1\) are displayed as 1.
	}
	\label{fig:by-ampl-and-loss-p-tot-div-sq}
\end{figure}

To demonstrate the advantage provided by the non-Gaussian SC state, we compare the achievable sensitivity with the one provided by the Gaussian squeezed states with the equivalent value of the squeeze factor, equal by the absolute value to the factor \(r\) used in the previous estimates, assuming the same quantum efficiency \(\eta\). It is shown in Appendix~\ref{app:sq-overlap}  that the corresponding error probability for detection of a given phase shift is equal to 

\begin{equation}\label{eq:p-sq-total-error}
	p_{\rm sq} = \text{erfc}\left(\delta_0\sqrt{\frac{\eta}{2\eta\left(s^2-1\right)+2}}\right) .
\end{equation}
In Fig.~\ref{fig:by-ampl-and-loss-p-tot-div-sq}, the corresponding ratio \(p_{\rm tot}/p_{\rm sq}\) is plotted as a function of $\alpha$ and $\eta$. For each pair $(\alpha, \eta)$, the optimal values of $\delta_0$, $r$ are used, as in Fig.~\ref{fig:by-ampl-and-loss-p-tot}. It can be seen from this plot that, in the case of small losses, the SC state could provide several times better sensitivity. 

In Appendix~\ref{app:squeeze-and-displacement-tolerances}, we analyze the tolerances of two central parameters: the antisqueezing parameter \(r\) and the displacement parameter \(\delta_0\). The tolerances are defined as the maximum independent deviations that keep the total discrimination error \(p_{\rm tot}\) within 10\% of its optimized minimum. For the experimentally feasible parameter range \(0.96\leq\eta\leq0.99 \) and \(1.5\leq\alpha\leq3.0\), the tolerances are as follows: \(\Delta \delta_0 \geq 0.06\) and \(\Delta r \geq 0.03\), the latter corresponding to about 0.3 dB of antisqueezing variation.

The required squeezing tolerance is compatible with the capabilities of state-of-the-art precision interferometers. For example, the Laser Interferometer Gravitational-Wave Observatory (LIGO) has demonstrated control of the squeezing level at the approximately 0.3 dB scale ~\cite{Tse_PRL_123_23_2019}. Using Eq.~\eqref{eq:delta-0-defenition}, the displacement tolerance \(\Delta \delta_0 = 0.06\) can be expressed as the corresponding interferometric phase-shift tolerance

\begin{equation}
	\Delta \phi = \frac{\Delta \delta_0}{\sqrt{N}} \approx \frac{0.06}{\sqrt{N}} .
\end{equation}

For \(r\geq2\), this phase tolerance becomes comparable to the shot-noise limit for squeezed states [see Eq.~\eqref{dphi_sqz}]. Consequently, realizing the predicted advantage requires a strongly phase-stabilized interferometric platform that can maintain reliable operation at the corresponding squeezing level.

\section{Conclusion} \label{sec:discussion}

We showed here that by using antisqueezed Schrödinger cat probe states in the non-Gaussian interferometric scheme proposed in Refs.~\cite{Gorshenin_LPL_21_6_2024, Gorshenin_JOSA_B_42_7_2025} it is possible to make this scheme significantly more tolerant to optical losses.

In our analysis, we assumed the use of the modern photon-number-resolving detectors and the maximum-likelihood decision rule. We performed numerical optimization of the antisqueezing factor assuming optimistic but experimentally achievable values of the Schrödinger cat amplitude $\alpha$ and the quantum efficiency of the setup $\eta$, and identified the area in these parameters space within which the resulting sensitivity overcomes the one provided by the Gaussian squeezed states. Note that this area corresponds to demanding but feasible values of \(\eta \gtrsim 0.96\) and SC state amplitudes \(1.5\leq\alpha\leq3\).

\acknowledgments

This work was supported by the Foundation for the Advancement of Theoretical Physics and Mathematics (``BASIS'' Grant No. 23-1-1-39-1). The authors would like to express deep gratitude to F. Ya. Khalili for his invaluable contributions to the discussions. 

\section*{Data Availability}
The data that support the findings of this article are openly available, embargo periods may apply. The code which was used for obtaining these results is provided in the GIT repository \cite{github_repo}. 

\appendix
\section{Two sequential photon losses}\label{app:two-seq-phot-loss}

Consider the following sequence: 
\begin{equation}
	\hat{\mathcal{U}} = \hat{\mathcal{L}}_2\hat{\mathcal{D}}(i\delta_0)\hat{\mathcal{L}}_1 \,,
\end{equation}
where $\mathcal{D}(i\delta_0)$ is the displacement operator \eqref{eq:displacement_oper_definition}, $\hat{\mathcal{L}}_i$, with $i=1,2$, are the evolution operators describing the optical losses as follows:
\begin{subequations}
	\begin{gather}
		\hat{\mathcal{L}}_i^\dag\hat{a}\hat{\mathcal{L}}_i
		= \sqrt{\eta_i}\hat{a} + \sqrt{1-\eta_i}\hat{z}_i \,, \label{L(a)}\\
		\hat{\mathcal{L}}_i^\dag\hat{z}_i\hat{\mathcal{L}}_i
		= \sqrt{\eta_i}\hat{z}_i - \sqrt{1-\eta_i}\hat{a} \,,
	\end{gather}
\end{subequations}
and $\hat{z}_{1,2}$ are the annihilation operators of the two vacuum modes. Using the unitarity of the operator $\hat{\mathcal{L}}_2$ and Eq.\,\eqref{L(a)}, we obtain that:
\begin{equation}
	\hat{\mathcal{U}}
	= \hat{\mathcal{L}}_2\hat{\mathcal{D}}(i\delta_0)\hat{\mathcal{L}}_2^\dag
	\hat{\mathcal{L}}_2\hat{\mathcal{L}}_1
	= \hat{\mathcal{D}}(i\delta_0\sqrt{\eta_2})\hat{\mathcal{D}}_{\rm loss}\hat{\mathcal{L}}
	\,,
\end{equation}
where
\begin{gather}
	\hat{\mathcal{D}}_{\rm loss} = e^{i\delta_0\sqrt{1-\eta_1}(\hat{z}_1 - \hat{z}_1^\dag)}
	\,,\,
	\hat{\mathcal{L}} = \hat{\mathcal{L}}_2\hat{\mathcal{L}}_1 \,.
\end{gather}

Note that
\begin{equation}
\begin{aligned}
	\hat{\mathcal{L}}^\dag\hat{a}\hat{\mathcal{L}} =
	\sqrt{\eta_2}(\sqrt{\eta_1}\hat{a} + \sqrt{1-\eta_1}\hat{z}_1) + \sqrt{1-\eta_2}\hat{z}_2
	=\\= \sqrt{\eta}\hat{a} + \sqrt{1-\eta}\hat{z} \,,
\end{aligned}
\end{equation}
where
\begin{gather}\label{eta_eff}
	\eta = \eta_1\eta_2 \,,\,
	\hat{z} = \frac{\sqrt{(1-\eta_1)\eta_2} + \sqrt{1-\eta_2}}{\sqrt{1-\eta_1\eta_2}}
\end{gather}
are the unified quantum efficiency and the corresponding effective noise; compare with Eq.~\eqref{L(a)}. Note also that $\hat{\mathcal{D}}_{\rm loss}$ is a unitary operator acting only in the first losses mode subspace and therefore vanishes when calculating the output state of the signal mode:

\begin{equation}
	\begin{aligned}
		\hat{\rho}_{\rm signal\,out}
		= \Tr_{\rm loss}(\hat{\mathcal{U}}
		\hat{\rho}_{\rm signal\,in}\otimes\hat{\rho}_{\rm loss}\hat{\mathcal{U}}^\dag)
		=\\=
		\Tr_{\rm loss}(
		\hat{\mathcal{D}}(i\delta_0\sqrt{\eta_2})\hat{\mathcal{L}}
		\hat{\rho}_{\rm signal\,in}\otimes\hat{\rho}_{\rm loss}
		\hat{\mathcal{L}}^\dag\hat{\mathcal{D}}^\dag(i\delta_0\sqrt{\eta_2})
		) .
	\end{aligned}
\end{equation}
where $\hat{\rho}_{\rm signal\,in}$ in and $\hat{\rho}_{\rm loss}$ are the initial states of the signal and the loss modes and $\Tr_{\rm loss}$ means tracing out the loss modes. 

This result corresponds to the action of the unified losses with the quantum efficiency \eqref{eta_eff}, followed by the lossless evolution of the signal mode, with the displacement factor scaled by $\sqrt{\eta_2}$.

\section{Derivation of the Wigner function after passing equivalent optical scheme}\label{app:wig-func-calc}

The corresponding Wigner function for the state \(\ket{\psi_{\rm sq,\,cat}}\) (see Eq.\,\eqref{eq:sq-sc-state-def}) is given by:
\begin{equation}\label{eq:cat-state-wigner-func-before-losses}
	\begin{aligned}
		W(x,p) = \frac{2}{\pi K}
		\Big\{
		\exp\Big(-2 \left(x s - \alpha\right)^2 - 2 \big(p/s\big)^2\Big)
		+ \\ +
		\exp\Big(-2\left(x s + \alpha\right)^2 -2 \big(p/s\big)^2\Big)
		+ \\ +
		2 \exp\Big(-2\left(xs\right)^2 - 2 \big(p/s\big)^2\Big) \cos \Big(4 \alpha p/s\Big)
		\Big\}\,,
	\end{aligned}
\end{equation}
where \(s \equiv e^r\).

Taking into account the input losses (see Ref.~\cite{Leonhardt_PRA_48_4598_1993}), we obtain the following Wigner function of the effective input state of the interferometer:

\begin{equation} \label{eq:wig-func-after-losses-ref}
	W^{(\eta)}(x,\,p) = \iint W(u,\,v) B(x- \sqrt{\eta} u,\,p- \sqrt{\eta} v,\,\eta) du dv \,,
\end{equation}
where
\begin{equation}
	B(x, p, \eta) = \frac{2}{\pi(1-\eta)} \exp\Big( -\frac{2(x^2 + p^2)}{1-\eta}\Big)
\end{equation}
is the ``Gaussian blurring filter''.

This Wigner function can be presented as follows:
\begin{equation}
	\begin{aligned} \label{eq:cat-state-wigner-func}
		W^{(\eta)}
		=
		F
		\Big(
		W_{\rm int}
		+
		W_+
		+
		W_-
		\Big) ,
	\end{aligned}
\end{equation}
where we introduce the ``bell'' terms \(W_{\pm} (x, p)\), corresponding to the two coherent-state components, and the interference term \(W_{\rm int} (x,p)\) of the Wigner function 

\begin{equation}\label{eq:bell_part_sc_wigner_prelude}
	W_{\pm} = \exp \Big(
	-A (x\pm\xi)^2 - B p^2
	\Big)
\end{equation}
\begin{equation}\label{eq:int_part_sc_wigner_prelude}
	W_{\rm int} = 2 \exp
	\Big(
	-C
	-B p^2
	-A x^2
	\Big)
	\cos
	\Big(
	D p
	\Big)
\end{equation}
where we introduce the effective SC state amplitude after loss, \(\xi = \frac{\alpha \sqrt{\eta}}{s}\). The non-negative coefficients \(A, B, C, D\) and \(F\) are presented in Eq.\,\eqref{eq:sc-state-wigner-func-pars}.

After applying the displacement operator with the displacement parameter \(\delta = \delta_0 \sqrt{\eta_2}\), we obtain the same functional form of the Wigner function [see Eq.~\eqref{eq:cat-state-wigner-func}], but with slightly modified bell and interference terms [see Eq.~\eqref{eq:int_part_sc_wigner} and Eq.~\eqref{eq:bell_part_sc_wigner}].

\section{
	Derivation of negative volume of antisqueezed SC state after losses [Eq.\,\eqref{eq:negative-volume-sc-state-sq-after-losses-approx}]} \label{app:negative-volume-of-sc-stae-derivation}

To obtain an analytical estimate of the Wigner negativity, we make the following approximations. First, we keep only the interference contribution Eq.~\eqref{eq:int_part_sc_wigner}, and drop the two Gaussian bell terms in Eq.~\eqref{eq:bell_part_sc_wigner}. We discuss the regime of validity of this simplification in the end of this appendix. With this approximation, the negativity volume becomes

\begin{equation}
	V_{\rm neg} = \frac{1}{2}\int_{-\infty}^\infty dx \int_{-\infty}^\infty dp \left(\big|W_{\rm int}(x,p) \big|-W_{\rm int}(x,p)\right)\,.
\end{equation}

Second, in the case of bright-enough SC state, the interference fringes in the Wigner function [see Eq.~\eqref{eq:int_part_sc_wigner}] are much finer than the width of the Gaussian envelope associated with the two lobes. Because \(W_{\rm int} \sim \exp(-B(y-\delta)^2)\cos(D(p-\delta))\) it oscillates rapidly across the momentum axis, generating alternating positive and negative fringes within a slightly changing envelope during one cosine period 

\begin{equation}
	W_{\rm int} \sim \cos(D(p-\delta)) \exp(-B(p-\delta)^2)\,,
\end{equation}

Accordingly, we restrict the integration domain to the negative-fringe regions where \(W_{\rm int}(x,p) < 0\), i.e., where \(\cos(D(p-\delta)) < 0\). In the case of bright-enough SC state, the cosine fringes are fast compared with the variation scale of the Gaussian envelope, so the integral over phase space can be evaluated by decomposing it into a sum over individual negative half-periods and by treating the Gaussian envelope as approximately constant within each half-period. Under this slowly varying envelope approximation, the integral becomes 

\begin{equation}
	\begin{aligned}
		V_{\rm neg} =
		2 \exp(-C) \int \exp(-A x^2) dx
		\sum_{k=-\infty}^{k=\infty}
		\exp(-B(p_k - \delta)^2)
		\cdot \\ \cdot
		\int_{p_k - \pi/(2D)}^{p_k + \pi/(2D)} \cos(D(p-\delta)) dy
		=\\=
		4 \vartheta_3( \exp(-((4 B \pi^2)/D^2)))\,,
	\end{aligned}
\end{equation}
where \(p_k = \delta + 2 \pi k/D,\,(k \in \mathbb{Z})\) denote the positions of the cosine minima.

Our simplified treatment rests on two assumptions. First, the interference contribution \(W_{\rm int}\), which carries the Wigner negativity, is well separated in phase space from the positive Gaussian bell terms \(W_{\pm}\). This negligible-overlap condition can be estimated using a standard three-sigma criterion where \(\sigma\) is the standard deviation of two terms of the Wigner function along the \(x\) axis 

\begin{equation}
	\sigma_{\rm bell, x} = \frac{1}{\sqrt{2 A}};\,\sigma_{\rm int, x} = \frac{1}{\sqrt{2 A}}\,.
\end{equation}

Substituting the parameters given in Eq.\,\eqref{eq:sc-state-wigner-func-pars} and simplifying yields the condition
\begin{equation}\label{eq:v-neg-approx-eq-applicability-condition}
	\frac{3 \alpha \sqrt{\eta} \sqrt{\left(\eta (s^2 - 1)+s^2\right)}}{s^2} < 1\,.
\end{equation}
The corresponding antisqueezing threshold \(s = s_{\rm max}\), defined by saturating the inequality, is also marked in  Fig.~\ref{fig:neg-volume-of-wigner-func}.

Our second approximation treats the slowly varying Gaussian envelope as approximately constant on the scale of a single interference fringe. We assess the regime of validity by comparing the exact expression in Eq.~\eqref{eq:int_part_sc_wigner} with the corresponding approximation at \(x=0\):

\begin{equation}
	f_{\rm fine}(p) \equiv
	2 e^{-C}
	e^{ -B(p-\delta)^2}
	\cos\Big(
	D (p-\delta)
	\Big)
\end{equation}
\begin{equation}
	\begin{aligned}
		f_{\rm approx}(p) \equiv
		2 e^{-C}
		\sum_{k = - \infty}^{k=\infty}
		e^{ -B(2 \pi k/D)^2}
		\cos\Big(
		D (p-\delta)
		\Big)
		\cdot\\\cdot
		\theta(p-\delta - 2 \pi k / D)
		\theta(-p+\delta + 2 \pi (k+1) / D)\,,
	\end{aligned}
\end{equation}
where \(\theta(x)\) is the Heaviside theta-function.

A convenient measure of the agreement is the inner product between \(f_{\rm fine}(p)\) and \(f_{\rm approx}(p) \), evaluated over the real line 

\begin{equation}
	I = \frac{\langle f_{\rm fine}, f_{\rm approx} \rangle
	}{
		\sqrt{
			\langle f_{\rm fine}, f_{\rm fine} \rangle
			\langle f_{\rm approx}, f_{\rm approx} \rangle
	}}
\end{equation}

where the denominator corresponds to the normalization constant. The inner product is defined as follows:

\begin{equation}
	\langle f, g \rangle
	\equiv
	\int_{-\infty}^{\infty} f_{\rm fine}(p) f_{\rm approx}(p) dp
\end{equation}

We report high-enough inner product value \(\langle f_{\rm fine}(p), f_{\rm approx}(p) \rangle\). For example, for \(\eta = 0.9\) the overlap is higher than 99.9\% for SC amplitudes \(\alpha > 6\) and antisqueezed in the range from 4 to 10 dB and losses. The lowest overlap value is equal to 99.8\% for small SC state amplitudes \(\alpha = 4\), which tell of an approximation correctness. This indicates that \(f_{\rm approx}(p)\) provides an accurate representation of \(f_{\rm fine}(p)\) and justifies the approximation used to estimate the Wigner-negativity volume.

\section{Derivation of Eq.\,\eqref{eq:prob-sc-loss-fine}}\label{app:phot-stat}

The Wigner function \(W_n(x,p)\) is the \(n\)-photon Fock state equal to
\begin{equation}
	W_{n}(x,p) =
	\frac{2}{\pi} (-1)^n e^{-2 (x^2 + p^2)}
	L_n(4 (x^2 + p^2))
\end{equation}
and
\begin{equation}
	L_n(z) = \sum_{k=0}^{k=n} \frac{(-1)^k}{k!} \frac{n!}{(n-k)! k!} z^k
\end{equation}
is the Laguerre polynomial.

Using the binomial expansion for powers of \(z = 4(x^2 + y^2)\), the Wigner function of the Fock state Laguerre polynomial can be written as
\begin{equation}
	L_n\Big(4(x^2+p^2)\Big) = \sum_{k=0}^{k=n} \sum_{m=0}^{m=k} \frac{4^k (-1)^k n!}{k!(n-k)!}
	\frac{1}{(k-m)! m!}
	x^{2m} p^{2(k-m)}\,.
\end{equation}

To calculate the photon-number probabilities according Eq.\,\eqref{eq:prob-fock-space-property}, we need to evaluate integrals of the form
\begin{equation}\label{eq:int_to_find_bell_overlapp_with_fock_state}
	\begin{aligned}
		I_{\rm int}^{(k,m)} = \int_{-\infty}^{\infty} \int_{-\infty}^{\infty}
		\cos \left(D (p - \delta)\right)
		\cdot \\ \cdot
		\exp\left( - A x^2 - B (p - \delta)^2\right) e^{-2 (x^2 + p^2)}
		x^{2m} p^{2(k-m)} dx dp \, ,
	\end{aligned}
\end{equation}
\begin{equation}\label{eq:int_to_find_int_overlapp_with_fock_state}
	\begin{aligned}
		I_{\rm bell, \pm}^{(k,m)} = \int_{-\infty}^{\infty} \int_{-\infty}^{\infty}
		\exp\left(- C -A(x \pm \xi)^2 - B (p - \delta)^2\right)
		\cdot \\ \cdot
		e^{-2 (x^2 + p^2)}
		x^{2m} p^{2(k-m)} dx dp \,.
	\end{aligned}
\end{equation}
As a result, we obtain:
\begin{equation}\label{eq:overlap-intermediate-res}
	p_n
	=
	4 \pi F
	\sum_{k=0}^{k=n} \sum_{m=0}^{m=k} \frac{4^k (-1)^k n!}{k! (n-k)!}
	\frac{I_{\rm int}^{(k,m)} + I_{\rm bell, +}^{(k,m)} + I_{\rm bell, -}^{(k,m)} }{(k-m)! m!}
	\,,
\end{equation}

Using WOLFRAM MATHEMATICA integrals \eqref{eq:int_to_find_bell_overlapp_with_fock_state} and \eqref{eq:int_to_find_int_overlapp_with_fock_state} could be calculated in the following form:
\begin{gather}
	\label{eq:I_bell_pm_transit}
	\begin{split}
		I_{\rm bell,\pm}^{(k,m)}
		=
		\frac{\Gamma\left(m+\frac{1}{2}\right)
			\Gamma\left(k-m+\frac{1}{2}\right) }{
			(A+2)^{m+\frac{1}{2}}
			(B+2)^{k-m+\frac{1}{2}}
		}
		e^{-A\xi ^2-B \delta ^2} \,
		\cdot\\\cdot
		{}_1F_1\left(m+\frac{1}{2};\frac{1}{2};\frac{A^2 \xi ^2}{A+2}\right)
		{}_1F_1\left(k-m+\frac{1}{2};\frac{1}{2};\frac{B^2 \delta
			^2}{B+2}\right) 
	\end{split}
\end{gather}

\begin{gather}
	\label{eq:I_int_transit}
	\begin{split}
		I_{\rm int}^{(k,m)}
		=
		2
		\frac{\Gamma\left(m+\frac{1}{2}\right)
			\Gamma\left(k-m+\frac{1}{2}\right) }{
			(A+2)^{m+\frac{1}{2}}
			(B+2)^{k-m+\frac{1}{2}}
		}
		e^{- B\delta ^2 - C}
		\cdot\\\cdot
		\Re\left(\,
		e^{-i D \delta } \,
		{}_1F_1\left(k-m+\frac{1}{2};\frac{1}{2};
		\frac{(i D+2 B \delta )^2}{4(B+2)}\right)
		\right)
	\end{split}
\end{gather}

The known properties could be used
\begin{equation}
\Gamma
\big(
\frac{1}{2} + n
\big)
_1F_1\Big(
n+\frac{1}{2};
\frac{1}{2};
z\Big)
=
\sqrt{\pi} e^z n! L_n^{-1/2} (-z)
\end{equation}
After this Eq.~\eqref{eq:I_bell_pm_transit} and Eq.~\eqref{eq:I_int_transit} slightly simplifies:
\begin{equation}\label{eq:I_bell_pm_simplified}
	\begin{aligned}
	I_{\rm bell,\pm}^{(k,m)}
	=
	\frac{\pi m! (k-m)!\exp\left(-\frac{2 A \xi^2}{A+2}-\frac{2 B \delta ^2}{B+2}\right)}{(A+2)^{m+\frac{1}{2}}
		(B+2)^{k-m+\frac{1}{2}} }
	\cdot\\\cdot
	L_m^{-\frac{1}{2}}\left(-\frac{A^2 \xi ^2}{A+2}\right)
	L_{k-m}^{-\frac{1}{2}}\left(-\frac{B^2 \delta ^2}{B+2}\right)
\end{aligned}
\end{equation}
\begin{equation}\label{eq:I_int_simplififed}
	\begin{aligned}
	I_{\rm int}^{(k,m)}
	=
	\frac{2 \sqrt{\pi } \Gamma
	\left(m+\frac{1}{2}\right) (k-m)! }{(B+2)^{k-m+\frac{1}{2}} (A+2)^{m+\frac{1}{2}} }
	\cdot\\\cdot
	\exp\left(-\frac{4(B+2) C+8 B \delta ^2+D^2}{4(B+2)}\right)
	\cdot\\\cdot
	\Re\left(\exp\left(i D \delta -\frac{i B D
		\delta }{B+2}\right)
	L_{k-m}^{-\frac{1}{2}}\left(\frac{(D+2 i B \delta )^2}{4 (B+2)}\right)\right)
\end{aligned}
\end{equation}
After substituting Eqs.~\eqref{eq:I_bell_pm_simplified} and \eqref{eq:I_int_simplififed} into Eq.~\eqref{eq:overlap-intermediate-res} and simplifying, we obtain Eq.~\eqref{eq:prob-sc-loss-fine}.

This expression can be simplified using Stirling’s approximation, which reduces numerical round-off errors and yields a more tractable form \cite{NIST:DLMF:Stirling-high-order}:
\begin{equation}
m! = \sqrt{2\pi m}\left(\frac{m}{e}\right)^m \exp\left(\frac{1}{12m} + \frac{1}{288m^2}\right)
\end{equation}

After this we obtain:
\begin{equation}
\frac{(2m)!}{(m!)^2}
\approx
\frac{2^{2m}}{\sqrt{\pi m}} S_m \, ,
\end{equation}
where was introduced factor \(S_m\) for brevity
\begin{equation}
\begin{aligned}
	S_m \equiv 
	\frac{\exp\big(\frac{1}{12 \cdot 2m} + \frac{1}{288 \cdot (2m)^2}\big)}{\Big(\exp\big(\frac{1}{12 \cdot m} + \frac{1}{288 \cdot m^2}\big)\Big)^2} = \exp\Big(-\frac{144 m + 7}{1152 m^2}\Big)
\end{aligned}
\end{equation}
According this:
\begin{equation}
\frac{\Gamma
	\big(
	\frac{1}{2} + m
	\big)}{m!}
=
\frac{(2m)!}{4^m (m!)^2} \sqrt{\pi} = \frac{S_m}{\sqrt{m}}
\end{equation}

After substituting into the Eq.~\eqref{eq:prob-sc-loss-fine} and reducing the terms, we obtain:
\begin{equation}\label{eq:prbability-sc-loss-aprox}
	\begin{aligned}
		p_n	=
		\sum_{k=0}^{k=n} \sum_{m=0}^{m=k}
		\frac{4 \pi n!}{k! (n-k)!}
		\frac{\pi F 2^{2 k+1}  (-1)^{k+n}}
		{(A+2)^{m+\frac{1}{2}}	(B+2)^{k-m+\frac{1}{2}} }
		\cdot\\\cdot
		\exp \left(-\frac{2 A \xi^2}{A+2}-\frac{D^2}{4 B+8}-\frac{2 B \delta
			^2}{B+2}-C\right)
		\cdot\\\cdot
		\Big( e^{\frac{D^2}{4 B+8}+C} L_m^{-\frac{1}{2}}\left(-\frac{A^2 \xi^2}{A+2}\right) L_{k-m}^{-\frac{1}{2}}\left(-\frac{B^2 \delta^2}{B+2}\right)
		+\\+
		e^{\frac{2 A \xi ^2}{A+2}}
		\frac{S_m}{\sqrt{m}}
		\Re\left(
		e^{\frac{2 i D \delta }{B+2}}
		L_{k-m}^{-\frac{1}{2}}
		\left(\frac{(D+2 i B \delta )^2}{4(B+2)}\right)
		\right)\Big)
	\end{aligned}
\end{equation}

We use this approximation in our numerical calculations when the parameter \(m>20\). We also note a practical limitation of the numerical procedure: the photon-number statistics can be computed reliably only up to a cutoff of \(n\leq 23\). For larger cutoffs, the calculation becomes numerically unstable and the accumulated round-off error becomes significant.

\section{
	Comparison of the photon-statistics calculation method proposed in Sec.~\ref{sec:phot-stat} with the combinatorial method}
\label{app:numerical-calc-comparison}

An alternative way to incorporate optical loss is to introduce conditional photon-counting probabilities. Below, we compare our method with this approach. Let the input (preloss) state be described by the density operator \(\rho_{\rm in}\) in the Fock basis. In the absence of loss, the probability of detecting \(n\) photons is the diagonal element \(p_n = (\rho_{\rm in})_{n,n}\). Optical loss transforms the photon-number distribution according to

\begin{equation}\label{eq:-losses-combinatorial-method}
	p_n^{(\eta)} = \sum_{m=n}^{\infty} P(n|m) p_m\,,
\end{equation}
where
\begin{equation}\label{eq:-losses-combinatorial-method-1}
	P(n|m) = \frac{m!}{n! (m-n)!} (1-\eta)^{m-n} \eta^n\,.
\end{equation}

To assess the accuracy of the Wigner-function-based photon-statistics method [Eq.~\eqref{eq:prbability-sc-loss-aprox}] relatively combinatorial method [Eq.~\eqref{eq:-losses-combinatorial-method-1}]. For this we define the following probability difference

\begin{equation}\label{eq:diff-probs-analytics-and-comb}
    \Delta = \sum_{n} |p_{n, \rm analytical} - p_{n, \rm cobinatoric}|
\end{equation}

where \(p_{n, \rm analytical}\) and \(p_{n, \rm cobinatoric}\) refers to Eq.~\eqref{eq:prbability-sc-loss-aprox} and Eq.~\eqref{eq:-losses-combinatorial-method-1} correspondingly. 

We report the small value of error of the proposed method of photon statistics calculation \(\Delta\) [see Eq.~\eqref{eq:diff-probs-analytics-and-comb}]. For example, consider the SC state with amplitude \(\alpha = 1.5\) and losses parameter \(\eta = 0.85\). For the band of squeezing \(0 \leq r \leq 1.2\) and displacement \(0\leq \delta_0 \leq 1.5\) the maximum error is \(\Delta = 0.002\) for the squeeze near zero. For the rest, the areas’ average error is \(\Delta \leq 0.001\).

\section{Derivation of Eq.\,\eqref{eq:p-sq-total-error}} \label{app:sq-overlap}

For comparison with the proposed protocol, we will compare it with an equally squeezed vacuum state. For this consideration, we consider that the vacuum state passes into considering the optical scheme (top panel) presented in Fig.~\ref{fig:scheme} and equivalent to it (bottom panel). Consequently, the vacuum state is squeezed, enforced to optical losses, and then displaced. For comparison we consider the homodyne measurement of momentum for the detection of a given phase shift. In that case, we aim to obtain the momentum probability representation of the resulting state.

The Wigner function of the displaced squeezed vacuum state is as follows:
\begin{equation}
	W_{\delta {\rm , sq}}(x,p) = \frac{2 e^{-2 s^2 x^2-\frac{2 (p-\delta_0)^2}{s^2}}}{\pi }
\end{equation}
where \(s = e^{-r}\) denotes the squeezed parameter along the momentum quadrature, rather than antisqueezing as in the SC state analysis. 

Similar to Sec.~\ref{sec:losses-on-sc-state-wigner-func}, after the ``Gaussian blurring filter'' the Wigner function is as follows: 
\begin{equation}
	W_{\delta {\rm , sq}}^{(\eta)}(x,p) = \frac{2 s \exp\left(
		-\frac{2 s^2 x^2}{\eta -\eta  s^2+s^2}-\frac{2 \left(p-\sqrt{\eta } \delta_0\right)^2}{\eta
			\left(s^2-1\right)+1}
		\right)}{\pi \sqrt{\left(\eta -(\eta -1) s^2\right) \left(\eta  \left(s^2-1\right)+1\right)}}
\end{equation}

For procedure of detection we will consider the homodyne measurement of momentum. It could be found easily from the Wigner function as follows: 
\begin{equation}
	\omega(p) = \frac{\sqrt{2} e^{-\frac{2 \left(y-\sqrt{\eta_2} \delta_0\right)^2}{\eta
				\left(s^2-1\right)+1}}}{\sqrt{\pi  \eta  \left(s^2-1\right)+\pi }}
\end{equation}

We aim to discriminate non-displaced and displaced vacuum states. Accordingly, it is reasonable to select a point between two Gaussian peaks along the momentum axis which is the border \(p_{\rm border}\) between displaced and non-displaced \(\delta_0 = 0\) states 

\begin{equation}
	p_{\rm border} = \sqrt{\eta_2} \delta_0 / 2
\end{equation}

Assume that the value \(\delta_0\) is positive. In the case of measuring momentum higher (lower) than \(p_{\rm border}\) we detect the presence (absence) of phase shift. Accordingly, these false-positive \(p_{\rm fp}^{\rm (sq)}\) and false-negative \(p_{\rm fn}^{\rm (sq)}\) fn errors could be introduced as follows: 

\begin{equation}
	p_{\rm fp}^{\rm (sq)}  = \int_{p=p_{\rm border}}^\infty \omega(p)  dp
	,\quad
	p_{\rm fn}^{\rm (sq)}  = \int_{-\infty}^{p=p_{\rm border}} \omega(p)\Big|_{\delta_0 = 0}  dp
\end{equation}
After calculating of integrals we obtain and assuming \(\eta_1 = 1\) and \(\eta_2 = \eta\):
\begin{equation}
	p_{\rm fp}^{\rm (sq)}
	=
	p_{\rm fn}^{\rm (sq)}
	=
	\frac{1}{2} \text{erfc}\left(\delta_0 \sqrt{\frac{\eta }{2 \eta  \left(s^2-1\right)+2}}\right)
\end{equation}
Total error for squeezed states \(p_{\rm sq}\equiv p_{\rm fp}^{\rm (sq)} +p_{\rm fn}^{\rm (sq)}\) is has form Eq.\,\eqref{eq:p-sq-total-error}.

\section{Tolerance requirements of antisqueezing and displacement parameters}\label{app:squeeze-and-displacement-tolerances}

In this appendix, we quantify the tolerances of two optimized parameters: the antisqueezing parameter \(r\) and the dimensionless displacement parameter  \(\delta_0\). The analysis is performed in the vicinity of the minimum of the total error probability, attained at \(\delta_{0,{\rm opt}}\) and \(r_{\rm opt}\). For each parameter, we define the tolerance as the deviation, \(\Delta r\) or \(\Delta \delta_0\), that increases the total detection error \(p_{\rm tot}\) by 10\% relative to its optimized minimum.

The antisqueezing tolerance \(\Delta r\) is determined as follows. At the minimum of \(p_{\rm tot}\), corresponding to \(\delta_{0,{\rm opt}}\) and \(r_{\rm opt}\), we first identify the optimal photon-number decision sets associated with the presence,  \(N_{\delta}\), and absence, \(N_0\), of the phase shift. These decision sets are then held fixed while the anti-squeezing parameter \(r\) is varied. The tolerance \(\Delta r\) is defined as the change in \(r\) for which the total error probability increases by 10\% relative to its optimized value:

\begin{equation}
	p_{\rm tot}(\delta_{0, {\rm opt}}, r_{\rm opt} + \Delta r) = 1.1 \times p_{\rm tot}(\delta_{0, {\rm opt}}, r_{\rm opt})
\end{equation}

If this condition admits multiple solutions for \(\Delta r\), we define the tolerance as the solution with the smallest absolute value, \(|\Delta r|\). Hereafter, \(\Delta r\) refers to this positive tolerance value.

The tolerance \(\Delta \delta_0\) for the displacement parameter is obtained using the same procedure. The condition is as follows:

\begin{equation}
	p_{\rm tot}(\delta_{0, {\rm opt}} + \Delta \delta_0, r_{\rm opt}) = 1.1 \times p_{\rm tot}(\delta_{0, {\rm opt}}, r_{\rm opt})
\end{equation}

\begin{figure}
	\centering
	\includegraphics[width=0.93\linewidth]{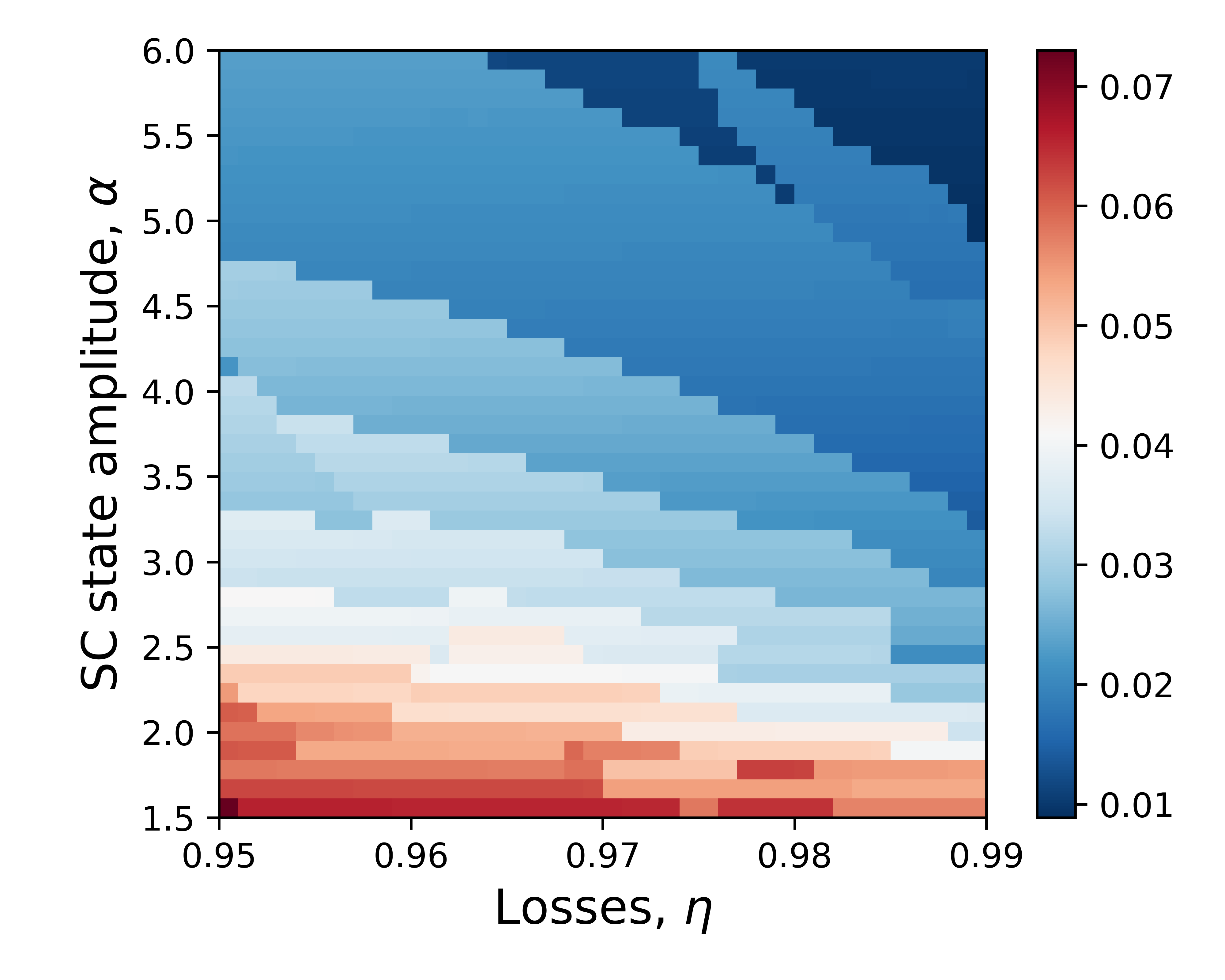}
	\caption{
Tolerance \(\Delta r\) of the anti-squeezing parameter, defined as the deviation from its optimal value that increases the total error by 10\% relative to the optimum (see Fig.~\ref{fig:by-ampl-and-loss-p-tot}), as a function of the SC-state amplitude \(\alpha\) and the quantum efficiency \(\eta\).
}
	\label{fig:squeeze-tolerances}
\end{figure}

Figure~\ref{fig:by-ampl-and-loss-r} shows that the optimized squeezing levels are experimentally attainable, but only with sufficiently precise tuning. The sensitivity to this tuning is quantified in Fig.~\ref{fig:squeeze-tolerances}. In the most optimistic experimentally accessible regime (\(\alpha = 6,\,\eta = 0.99\)), the allowed deviation \(\Delta r\) corresponds to a squeezing variation of only about \(\Delta r = 0.01\) which corresponds 0.1 dB before \(p_{\rm tot}\) increases by 10\%. For the more realistic parameter range (\(\eta \geq 0.96\) and \(1.5\leq \alpha \leq 3\)), the tolerance relaxes to \(\Delta r \simeq 0.03 - 0.07\) which corresponds 0.3--0.7 dB, while still maintaining the same 10\% error-increase criterion.

\begin{figure}
	\centering
	\includegraphics[width=0.93\linewidth]{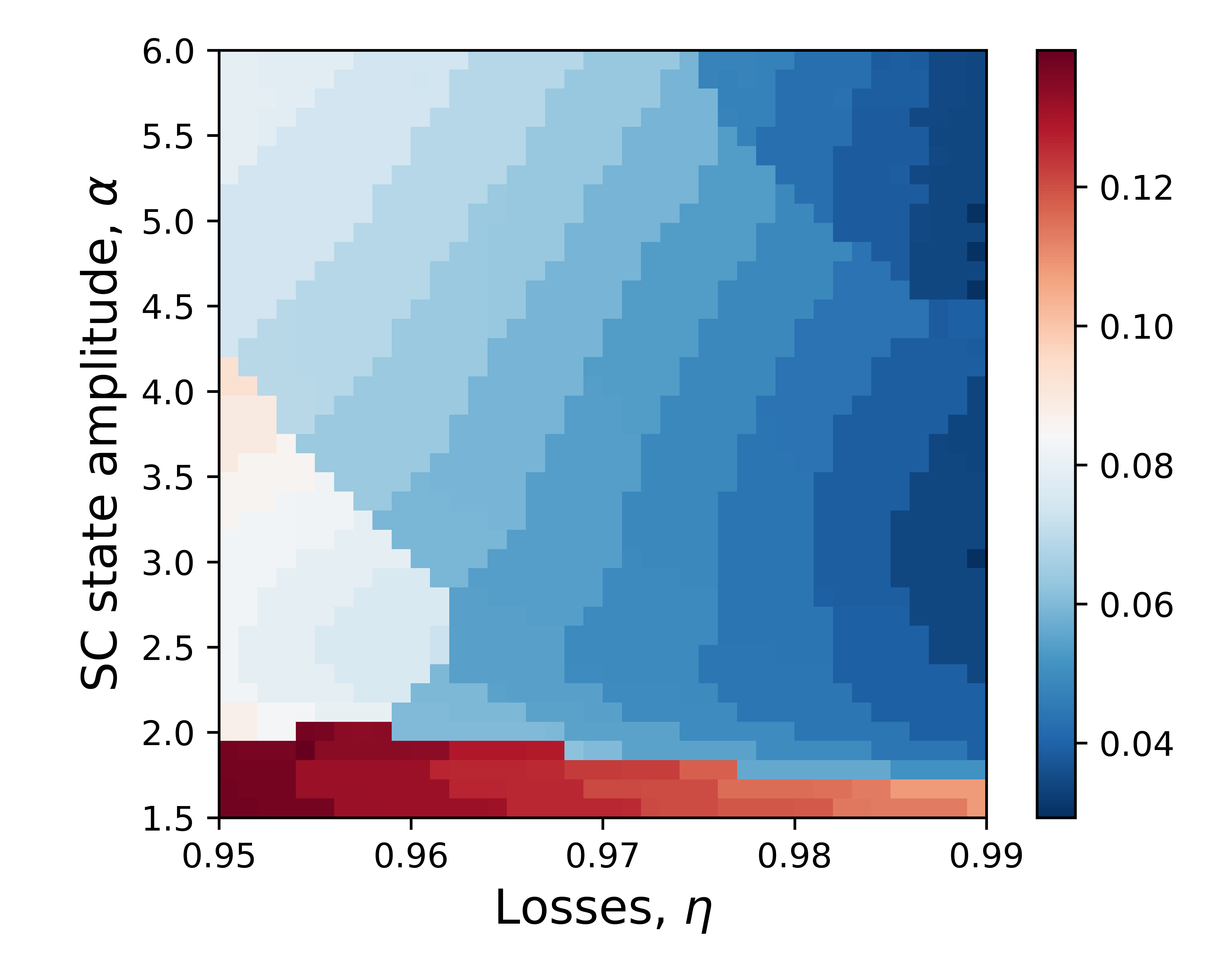}
	\caption{
		Tolerance \(\Delta \delta_0\) of the displacement parameter, defined as the deviation from its optimal value that increases the total error by 10\% relative to the optimum (see Fig.~\ref{fig:by-ampl-and-loss-p-tot}), as a function of the SC-state amplitude \(\alpha\) and the quantum efficiency \(\eta\).
		}
	\label{fig:displacement-tolerances}
\end{figure}

Finally, we consider the robustness of the anti-squeezed SC state relative to the squeezed Gaussian state under phase-shift deviations. Using the same 10\% increase in \(p_{\rm tot}\) as the tolerance criterion, we find that the required displacement accuracy remains experimentally feasible. As shown in Fig.~\ref{fig:displacement-tolerances}, the allowed deviation corresponds to \(\Delta \delta_0 \simeq 0.06 - 0.12\) in the dimensionless displacement parameter.

\end{document}